\documentclass[]{sagej}
\usepackage[]{graphicx}
\usepackage[]{color}
\makeatletter
\def\maxwidth{ %
  \ifdim\Gin@nat@width>\linewidth
    \linewidth
  \else
    \Gin@nat@width
  \fi
}
\makeatother

\definecolor{fgcolor}{rgb}{0.345, 0.345, 0.345}

\usepackage{framed}
\makeatletter
 {\par\unskip\endMakeFramed%
 \at@end@of@kframe}
\makeatother

\definecolor{shadecolor}{rgb}{.97, .97, .97}
\definecolor{messagecolor}{rgb}{0, 0, 0}
\definecolor{warningcolor}{rgb}{1, 0, 1}
\definecolor{errorcolor}{rgb}{1, 0, 0}

\usepackage{alltt}

\usepackage{epstopdf}
\usepackage[caption=false]{subfig}

\theoremstyle{plain}

\usepackage{hyperref}

\theoremstyle{definition}

\theoremstyle{remark}

\usepackage[utf8]{inputenc}
\usepackage{wrapfig}
\usepackage{longtable}
\usepackage{appendix}
\usepackage{adjustbox}

\usepackage[T1]{fontenc}

\usepackage{graphicx}
\usepackage{enumerate}

\usepackage[american]{babel}
\usepackage{csquotes}

\usepackage[numbers]{natbib}
\def\citepos#1{\citeauthor{#1}'s (\citeyear{#1})}

\usepackage{relsize}
\IfFileExists{upquote.sty}{\usepackage{upquote}}{}
\begin{document}


\title{Countering Underproduction of Peer Produced Goods}

\author{Kaylea Champion and Benjamin Mako Hill \\
 {\smaller \smaller Department of Communication, University of Washington, Seattle, Washington}
}

\begin{abstract}
Peer produced goods such as online knowledge bases and free/libre open source software rely on contributors who often choose their tasks regardless of consumer needs. These goods are susceptible to \textit{underproduction}: when popular goods are relatively low quality. Although underproduction is a common feature of peer production, very little is known about how to counteract it. We use a detailed longitudinal dataset from English Wikipedia to show that more experienced contributors---including those who contribute without an account---tend to contribute to underproduced goods. A within-person analysis shows that contributors' efforts shift toward underproduced goods over time. These findings illustrate the value of retaining contributors in peer production, including those contributing without accounts, as a means to counter underproduction.
\end{abstract}


\maketitle


\section{Introduction}
\label{sec:intro}

Peer production is a collaborative technology-assisted process in which participants select and perform tasks in a self-directed way, then combine their work with others \citep{benkler_wealth_2006}. Peer production plays a critical role in today's information ecosystem. From web servers to programming languages, Internet infrastructure is largely peer produced \citep{eghbal_roads_2016}. Much of the content we consume online is also peer produced. Peer produced websites such as Wikipedia, Reddit, and Fandom are among the most visited sites on the Internet.\footnote{\url{https://perma.cc/K5VM-V99E}} 
Peer produced content includes map data, search results, digital assistants responses, AI training data, and more \citep{mcmahon_substantial_2017}. 

How do producers' choices of what to make correspond to what consumers want to use? In a market-driven system, supply and demand are aligned via price. But in commons-based peer production, no such signal is available. This can lead to \textit{underproduction}: the production of high-interest but low-quality goods \citep{warncke-wang_misalignment_2015}.

This project seeks to understand the association between contributor experience, account use, and task selection. We first examine what is known about how tasks are selected to provide the rationale for several hypotheses. We then describe the setting, data, measures, and analytical plan. Next, we share empirical results. Finally, we discuss the limitations and implications of our results before concluding.  

\section{Background}
\label{sec:background}

\subsection{Commons-based Peer Production and Underproduction}
\label{subsec:ppInNPS}

Commons-based peer production is a term coined by Yochai Benkler to describe an emerging form of cooperative production made possible by new communication technology. Tasks are self-selected and combined through technology, resulting in information public goods. Examples of peer production include free/libre open source software (FLOSS) projects like GNU/Linux and Apache, Wikipedia, and OpenStreetMap (OSM).
\citet{benkler_wealth_2006} argues that because individuals know their interests, knowledge, availability, and skills, peer production creates the potential for diminished separation between producers and consumers and that self-assignment to tasks supports efficient matching between contributors and tasks. 

One way to assess the success of a peer production project is to examine how well it meets the needs of its users. 
\citet{warncke-wang_misalignment_2015} found that over 40\% of views to English Wikipedia were to articles that were lower quality than one might expect given their popularity.  
Countries, religions, LGBT topics, psychology, pop and rock music, the Internet and technology, comedy, and science fiction were found to be disproportionately affected. The misalignment of quality and public interest in these topics is concerning because of their relevance to public affairs and modern culture.

To quantify misalignment, \citet{warncke-wang_misalignment_2015} propose that if a system is aligned, goods in the highest demand should be the highest quality while those in the lowest demand should be the lowest quality. When quality is low relative to demand, goods are described as \emph{underproduced}. When quality is high relative to demand, they are \emph{overproduced}. Although overproduction may not be harmful beyond the potential for wasted effort, underproduction can impact public knowledge goods and digital infrastructure \citep{champion_underproduction_2021, eghbal_roads_2016}.

\subsection{Motivation, Experience, and Task Selection}
\label{subsec:experience_and_task}

To understand the sources of underproduction, we must consider why people participate in peer production. Studies have observed a range of motivations.
Some are motivated to create public goods \citep{budhathoki_motivation_2013}, to address social issues \citep{march_wikipedia_2020}, to help other community members \citep{wu_empirical_2007},   
to enhance their own use of a public good \citep{krishnamurthy_acceptance_2014, meng_commons/commodity:_2013}, to enhance their reputation or qualifications \citep{silva_google_2020,xu_empirical_2015,oreg_exploring_2008}, as an experience of self-efficacy \citep{yang_motivations_2010}, as a professional responsibility \citep{faric_motivations_2014}, for personal enjoyment and learning \citep{faric_motivations_2014}, out of reciprocity \citep{xu_empirical_2015}, for a class assignment \citep{mcdowell_wikipedia_2022,xing_editing_2020,coelho_why_2018,konieczny_wikis_2012}, or as part of their job \citep{germonprez_rising_2019}. The results of this body of literature suggest that contributor motivation in peer production tends to be complex and multidimensional and primarily, although not exclusively, intrinsic \citep{benkler_peer_2015, rafaeli_online_2008, belenzon_motivation_2015, kuznetsov_motivations_2006}.

To the extent that peer production contributors do so because they are told to (e.g., by an employer or instructor in a class), extrinsic rewards and others' interests may be primary motivators.
For example, contributing to Wikipedia is an increasingly common assignment in college classes where instructors are driven by concerns about content gaps or their commitment to public goods
\citep{mcdowell_wikipedia_2022, xing_editing_2020, konieczny_wikis_2012}, or by a desire to have students engage with their coursework in public \citep{gallagher_teaching_2019}. Firms may encourage their employees to contribute to FLOSS to enhance the firm's use of the public good or to build their reputation \citep{germonprez_rising_2019}.

Nonprofit organizations also encourage contributions to peer production projects as part of their missions. Examples include the Missing Maps project by the Red Cross and Doctors Without Borders encouraging contributions to OSM \citep{herfort_spatio-temporal_2023}, partnerships encouraging contributions to FLOSS projects targeting climate change,\footnote{e.g., Climage Triage \url{https://perma.cc/9R8F-QQYY}} and galleries, libraries and museums partnering with Wikipedia \citep{karczewska_glam_2023}. Some non-profits seek to counter content and participation gaps in peer production. For example, WikiEdu supports instructors using Wikipedia in their course assignments to close content gaps \citep{wilfahrt_improving_2022}. The Outreachy initiative aims to increase diversity in open source \citep{ackermann_future_2023}.

Within this complex motivational landscape, contributor task selection varies across users and within users over time. In their examination of OSM contributors, \citet{budhathoki_motivation_2013} reported that high-volume contributors were likelier to seek community recognition, while low-volume mappers reported wanting to contribute to a free and open project. 
In contrast, \citet{shah_motivation_2006} showed that FLOSS contributors describe their first contributions as directly related to personal skills, needs, and priorities, while longer-term participants report working for the good of the project.
Likewise, \citet{bryant_becoming_2005} found that while Wikipedia editors reported participating within their areas of expertise initially, they sought to build the Wikipedia community and serve the public good over time. \citet{restivo_no_2014} found that the highest-volume contributors produce more after receiving informal social rewards, while lower-volume contributors did not and may even respond negatively to the same awards.   

Previous work offers some additional clues to the relationship between experience and choosing to contribute to underproduced goods. New contributors initially select tasks that match their immediate needs, areas of knowledge, and skills \citep{preece_reader--leader_2009,bryant_becoming_2005}. 
Because people are interested in similar things, it seems likely that these areas of knowledge will, on average, be associated with high-interest subjects.
On the other hand, contributors faced with high-quality artifacts may not see where their assistance is needed \citep{preece_reader--leader_2009, bryant_becoming_2005}.
However, as they accumulate experience, we expect contributors' skills to increase.

Although this literature points in conflicting directions, prior work has emphasized that skilled participants incorporate the public's desire for information as a component of their task selection. As a result, we expect to find that the most experienced contributors are more likely to engage in the improvement of underproduced goods:
\textit{H1: individuals with less experience will contribute to less underproduced goods than individuals with more experience.}

In that experienced contributors may seek recognition for their work, social factors may act to disrupt the trend we describe in {H1} \citep{oreg_exploring_2008}. 
Once a participant is involved in a project, their ongoing participation may be encouraged by a desire for status \citep{willer_status_2009}.
To distinguish the extent to which social rewards drive task selection, we take advantage of the fact that receiving in-group social rewards partially depends on identifiability.
Collaborators may seek to direct their responses to an attributed person. Previous research suggests that creating an account may be driven by a desire to
obtain feedback or recognition \citep{forte_privacy_2017}. By contrast, \citet{belenzon_motivation_2015} argued that anonymous contributors to FLOSS projects were more motivated by a desire to contribute to the public interests, as measured by the fact that they were more likely to contribute to projects operated by a nonprofit than a for-profit entity, and those serving end users (rather than other developers).
Because contributing without an account makes social rewards more difficult, we propose that \textit{H2: contributing without an account is associated with more underproduced goods than contributing with an account.} 


Finally, we consider two competing explanations for {H1} and {H2}. An association between task selection and experience could be explained by contributor attrition (i.e., those who go on to make many contributions have always differed from those who make only a few) or by shifting motivations (i.e., persistence causes changes in task selection). In other words, if we find support for H1 and H2, is it due to differences caused by users being ``born'' or ``made''?
Support for the ``born'' explanation comes from past work that has found that individuals who go on to contribute at high volume seem to do so from the start \citep[e.g.,][]{panciera_wikipedians_2009}. On the other hand, contributors' motivation appears to shift over time \citep[e.g.,][]{shah_motivation_2006,bryant_becoming_2005}. Moreover, peer production projects have successfully invested in efforts to socialize and retain newcomers \citep[e.g.,][]{tan_first_2020, morgan_evaluating_2018}, suggesting that contributors can also be ``made'', at least to some extent.
Given this evidence of motivational shifts, we hypothesize within-person change as \textit{H3A: an individual will shift toward underproduced goods as they accumulate experience}, and \textit{H3B: an individual not using an account will shift toward underproduced goods at lower experience levels than one using an account.} Our reasoning for {H3B} is that because non-account-users are less likely to receive social rewards, they are even more likely to be motivated by the public interest.

\section{Research Design}
\label{sec:design}

\subsection{Empirical Setting}
\label{subsec:setting}
This study is conducted in the context of Wikipedia, the Wikimedia Foundation (WMF) website. 
Wikipedia is one of the most popular websites in the world and is used as reference material, a fact-checking source, and an input to machine learning and AI systems \citep{mcmahon_substantial_2017}. 
Wikipedia projects in 326 different language editions received 29 billion page views in January 2024.\footnote{
WMF Statistics: \url{https://perma.cc/H3NT-RP9N}, Wikipedia page about Wikipedia:
\url{https://perma.cc/RNC3-4SLN}} The largest Wikipedia language edition, English, has more than 6.7 million articles.
With some exceptions, clicking on an `Edit' tab on every Wikipedia page creates an interface for revising the existing text. For most language editions, changes are immediately visible without review.


\subsection{Data}
\label{subsec:sample}

Our unit of analysis is the \textit{revision}, and we use the complete revision history of Wikipedia, made available by the WMF, from its inception in January 2001 through July 2021. The complete revision history of Wikipedia contains a wide range of content, including discussion and personal pages. 

Our hypotheses concern the relationship between experience and having an account with article underproduction, a function of viewership and quality. To test our hypotheses, we filtered the revision population in multiple ways before drawing our sample. We excluded work done by bots, revisions to non-article pages, vandalism and its removal, and purely administrative revisions. 
Because no viewership data are available before December 2007, we only considered contributions after December 2007.
Details on these filters and the total revisions after each step are reported in our online supplement.

In Wikipedia, a small fraction of contributors make a large portion of all contributions. To capture variation based on editor experience, we drew two samples. For the \textit{revision sample} used to test {H1} and {H2}, we first stratified contributions by their ordinal position into increasingly large ``buckets'' increasing in size by $2^x$ (i.e., 1, 2, 4, 8, and so on).
We then drew equal-sized random samples from each bucket, oversampling where the data was thinner, for a total of 192,672 revisions. The inverse of the resulting sampling proportions was used as weights during subsequent calculations \citep{tracy_adjusting_2014}.


The second sample is a \textit{within-person sample}, designed to extract the contribution history of editors who differed in their ultimate contribution level. 
To build this sample, we stratified editors based on their total contributions through July 2021, again by creating increasingly large ($2^x$) buckets. We then drew a random sample of editors from each. Again, we oversampled where data was thinner and retained proportions as weights. After selecting 
editors, we gathered all article contributions made by the selected editors for a total of 42,602,912 revisions. This approach gives us the contribution history of a diverse range of editors: those who went on to contribute a great deal and those who only contributed once or twice.

We have made our data and code available at \url{https://doi.org/10.7910/DVN/UDQT6E}.

%

\subsection{Measures}
\label{subsec:variables}

Our dependent variable is \textit{underproduction factor}. To construct this measure, we drew from elements of both \citet{champion_underproduction_2021} 
and \citet{warncke-wang_misalignment_2015} and calculated the negative log of the ratio between quality rank and popularity rank. Both ranks were assigned such that high quality and high popularity receive high ranks, while low quality and low popularity receive low ones. A perfectly aligned article---one with the same popularity and quality rank---would have an underproduction factor of 0 since log(1) is 0. An overproduced article of high quality and low popularity would have a negative underproduction factor. An underproduced article of low quality and high popularity would have a positive underproduction factor.

We count views at the monthly level using data released by the WMF.
We calculate quality at the monthly level using the ORES quality measure. ORES is a machine learning-based quality measure that reflects the structural characteristics of articles, such as the presence of references, section headings, and inter-wiki links \citep{halfaker_ores:_2016}. ORES is a continuous measure with separate predictions for six article quality levels. We treat these levels as equally spaced to create a single quality score between 0 and 5. More details are provided in the supplement.

We operationalize editor experience as \textit{revision count}: the number of revisions the editor has made before and including the current revision. 
Therefore, before conducting the filtering described in the data and sample section, we calculated each revision's ordinal position in its contributor's edit history, ignoring any reverted contributions.
To consider the influence of the potential for receiving social rewards, we also introduced a variable reflecting whether contributions were made by users who were not using accounts. Because these users are identified by IP address in Wikipedia, we call this measure \textit{IP-based}.

\subsection{Analytical Plan}
\label{subsec:analyticalprocess}

We employ a hierarchical multiple regression model to test our hypotheses {H1} and {H2}. We use \textit{revision count (logged)}, \textit{revision count (logged) squared}, and \textit{IP-based} as predictors and a random intercept term for users using the \textit{lmer} package in R and a `raw' specification for our polynomials.

Although our hierarchical models attempt to correct for repeated measures of users, this approach does not account for the fact that experienced users might edit different articles than less experienced users in ways that reflect underlying differences in the types of users who go on to become less or more experienced \citep{panciera_wikipedians_2009}. To answer {H3A} and {H3B}, we fit a second group of models with user-level fixed effects using the \textit{felm} package in R and our within-person sample. Because we have no within-person variation in account use, we cannot include \textit{IP-based} in the within-person model.
Due to heteroskedasticity, we use robust standard errors in all our models.

\subsection{Ethics}
This study was conducted entirely using publicly available data published by WMF and does not involve any interaction or intervention with human subjects. The IRB at our institution has reviewed this type of research using these data and determined it is not human-subject research. WMF fully anonymized article view data before release.

\section{Results}
\label{sec:results}

\begin{figure}
\centering
\includegraphics[width=0.6\textwidth]{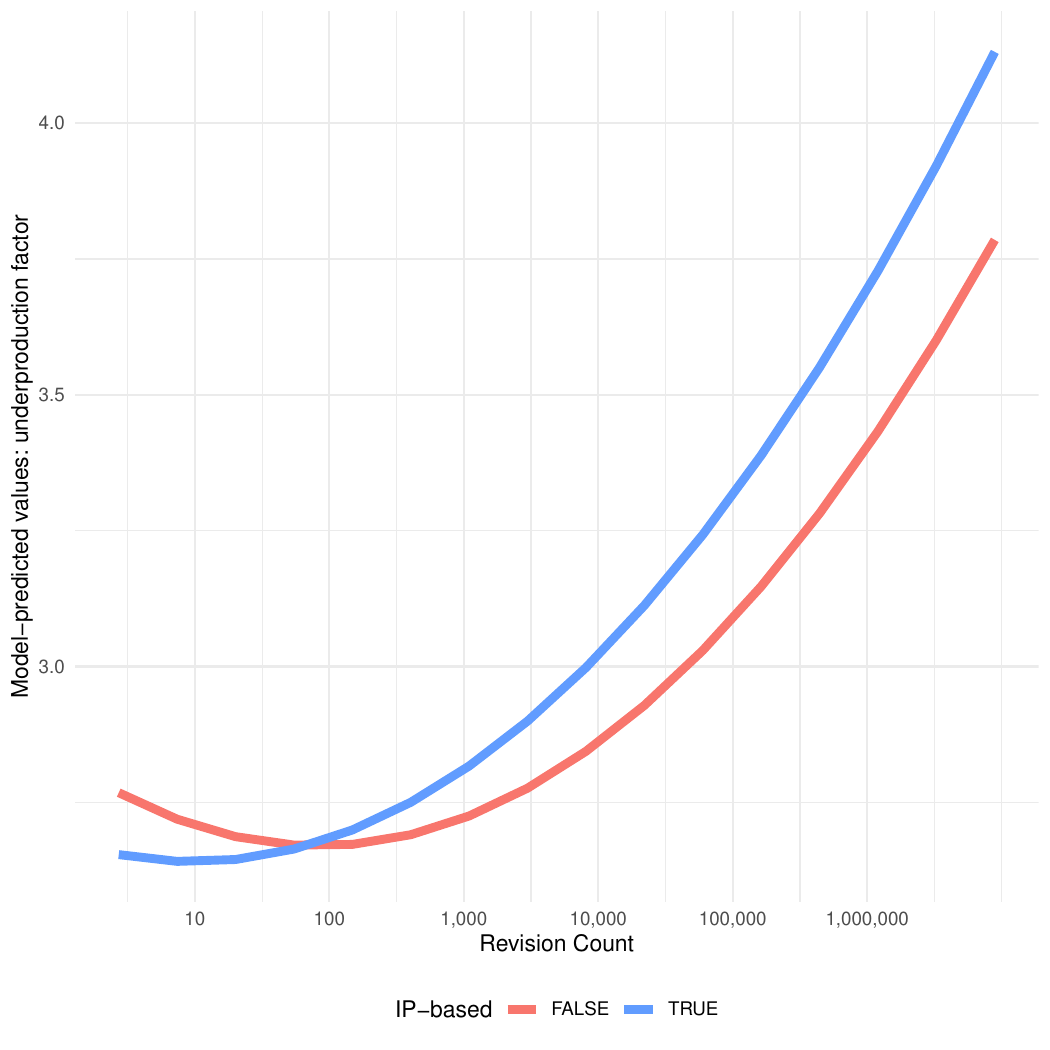}
\caption{The marginal effect of having higher experience on the average alignment of an article selected for editing. Increasing values indicate increased levels of underproduction, i.e. low-quality but highly-viewed topics.}
\label{fig:AlignMargEffPoly}
\end{figure}

The results of our polynomial model evaluating the relationship between underproduction and experience ({H1}) and between underproduction and having an account ({H2}) are shown in Table \ref{tab:betweenAlignPoly} and visualized in a marginal effects plot in Figure \ref{fig:AlignMargEffPoly}. 

\begin{table}

\begin{tabular}{l c c}
\hline
 & Linear Model & Polynomial Model \\
\hline
Intercept                                     & $2.5604$              & $2.8029$              \\        
                                                & $ [ 2.5358;  2.5851]$ & $ [ 2.7682;  2.8377]$ \\        
Revision Count (ln)         & $0.0287$              &   $-0.0738$             \\        
                                                & $ [ 0.0250;  0.0323]$ & $ [-0.0847; -0.0628]$ \\        
Revision Count (ln$^2$)          &                       & $0.0083$              \\        
                                                &                       & $ [ 0.0075;  0.0092]$ \\        
Editor was IP-based                                        & $0.0609$              & $-0.1511$             \\        
                                                & $ [ 0.0303;  0.0916]$ & $ [-0.1920; -0.1103]$ \\        
Revision Count (ln) and IP-based  & $-0.0198$        & $0.0376$              \\        
                                                & $ [-0.0272; -0.0123]$ & $ [ 0.0165;  0.0586]$ \\        
Revision Count (ln$^2$) and IP-based  &                       & $-0.0004$             \\        
                                                &                       & $ [-0.0035;  0.0027]$ \\        
\hline
AIC                                             & $656075$         & $655675$         \\        
BIC                                             & $656136$         & $655756$         \\        
Log Likelihood                                  & $-328032$        & $-327829$        \\        
Num. obs.                                       & $192672$              & $192672$              \\        
Num. groups: editor\_id\_or\_ip                 & $89445$               & $89445$               \\        
Var: editor\_id\_or\_ip (Intercept)             & $0.4476$              & $0.4510$              \\        
Var: Residual                                   & $164.7730$            & $164.2004$            \\        
\hline
\end{tabular}

\caption{Results from our revision-level hierarchical model of average alignment level of articles selected for editing. Bracketed values indicate a 95\% confidence interval.}
\label{tab:betweenAlignPoly}
\end{table}

In the linear model for {H1}, we find that an increase in one log unit of experience is associated with a 0.0287 increase in the underproduction factor of the article selected for contribution (see Table \ref{tab:betweenAlignPoly}). The quadratic specification of our model is shown alongside the linear specification and is a substantially better fit for the data ($\delta2LL\approx 200$; $\chi^2= 405$; $df=2$; $p < .0001 $). The parameter estimates for all terms in both models are statistically significant, except for the interaction of revision count and IP-based. The polynomial model describes U-shaped curves with an inflection point at about 150 revisions for those contributing with an account (see Figure \ref{fig:AlignMargEffPoly}). 
These results provide some support for {H1}: contributions by low-experience individuals are to less underproduced articles than those from users with high experience levels. However, as we see from the polynomial model, the relationship is U-shaped. We also note that the highly skewed distribution of contributions is such that most contributors do not make more than a small number of contributions. 

\begin{figure}
    \centering
    \includegraphics[width=.7\textwidth]{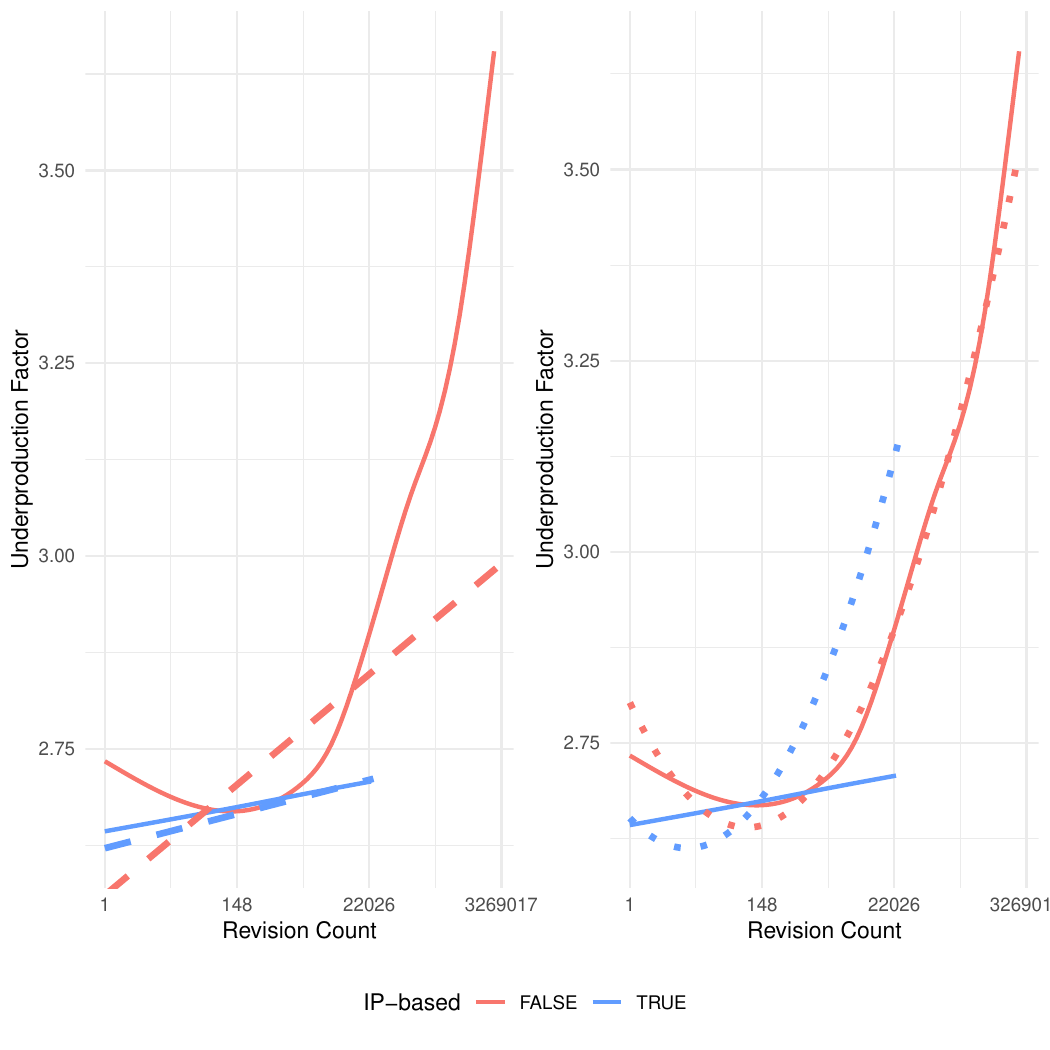}
    \caption{The left pane shows a GAM smoothed line fit to a 10\% sample of the data from the random sample with the result of the linear model superimposed as a dashed line; the right pane shows the same GAM line with the result of the polynomial model superimposed as a dotted line. Note the log-scaled $x$-axis.}
    \label{fig:polyLinearPanes}
\end{figure}

For {H2}, our linear model shows that contributing using an IP address is associated with a 0.0609 unit increase in the underproduction factor of the selected article. An increase in one log unit of experience is associated with a difference of -0.0198 in the same outcome. The quadratic specification of our model shows the opposite: the parameter estimate of the main effect associated with using an IP address is a -0.1511 difference in underproduction factor, with a positive interaction term (revision count and IP-based) of 0.0376. The squared interaction term (revision count squared and IP-based) is both small in magnitude and not statistically significant. 

To help understand these results, we construct Figure \ref{fig:polyLinearPanes}, which shows a nonparametric GAM smoothed line fit to a 10\% sample of the same data used to fit the models alongside the predicted values from both the linear and polynomial models. 
While the quadratic model is a better fit for the data, the linear model is more consistent with the behavior of IP-based contributors as shown in our nonparametric data visualization.
If we use the linear model as the standard for evaluating our hypothesis about IP-based editors, we find that although they shift towards underproduced materials as they accumulate experience, they do not do so at greater levels than those contributing with an account. 
Overall, these results contradict our proposal in {H2}. 


\begin{table}

\adjustbox{max width=\textwidth}{

\begin{tabular}{l c c c c}
\hline
 & Within-Person: With Account & IP-based & With Account, Quadratic & IP-based, Quadratic \\
\hline
Revision Count (ln)                       & $0.0808$            & $0.0219$                        & $-0.0193$             & $-0.0253$             \\
                                        & $ [0.0803; 0.0812]$ & $ [0.0174; 0.0264]$ & $ [-0.0205; -0.0180]$ & $ [-0.0342; -0.0164]$ \\
Revision Count (ln$^2$)                      &                     &                                  & $0.0077$              & $0.0087$              \\
                                       &                     &                     & $ [ 0.0076;  0.0078]$ & $ [ 0.0073;  0.0101]$ \\
\hline
Num. obs.                                        & $42170545$          & $432367$             & $42170545$            & $432367$              \\
R$^2$ (full model)                               & $0.1830$            & $0.3575$                   & $0.1835$              & $0.3578$              \\
R$^2$ (proj model)                               & $0.0030$            & $0.0002$            & $0.0036$              & $0.0006$              \\
Num. groups: editor\_id\_or\_ip        & $9054$              & $4290$                & $9054$                & $4290$                \\
\hline
\end{tabular}

}

\caption{Results of cross-sectional panel modeling to examine within-person change in the alignment of articles selected. Bracketed values indicate a 95\% confidence interval using robust standard errors. The projected $R^2$ indicates modeling this data with random effects for experience level but without fixed effects for the individual would result in poor model fit.}
\label{tab:withinAlign}
\end{table}

{H3A} and {H3B} relate to within-person change. The results of our model testing both parts of {H3} are presented in Table \ref{tab:withinAlign}. In {H3A}, we proposed a within-person shift toward underproduced goods as contributors increase in experience. 
All models support {H3A}. We observe a positive coefficient in the linear specification, suggesting an upward trend toward increasingly underproduced articles as individuals accumulate experience. In the quadratic specification, we observe a positive coefficient associated with the second-order term, suggesting a U-shaped relationship such that at higher revision counts, contributors select underproduced articles. We see mixed results for {H3B} related to those contributing without an account. When fit with a linear specification, the magnitude of the effect of experience is smaller for those contributing without an account (0.0219) than the coefficient associated with contributing with an account (0.0808). However, in the quadratic model, the coefficient in the squared term is slightly larger for those contributing without an account (0.0087) than those contributing with an account (0.0077). 

\begin{figure}
    \centering
    \includegraphics[width=.6\textwidth]{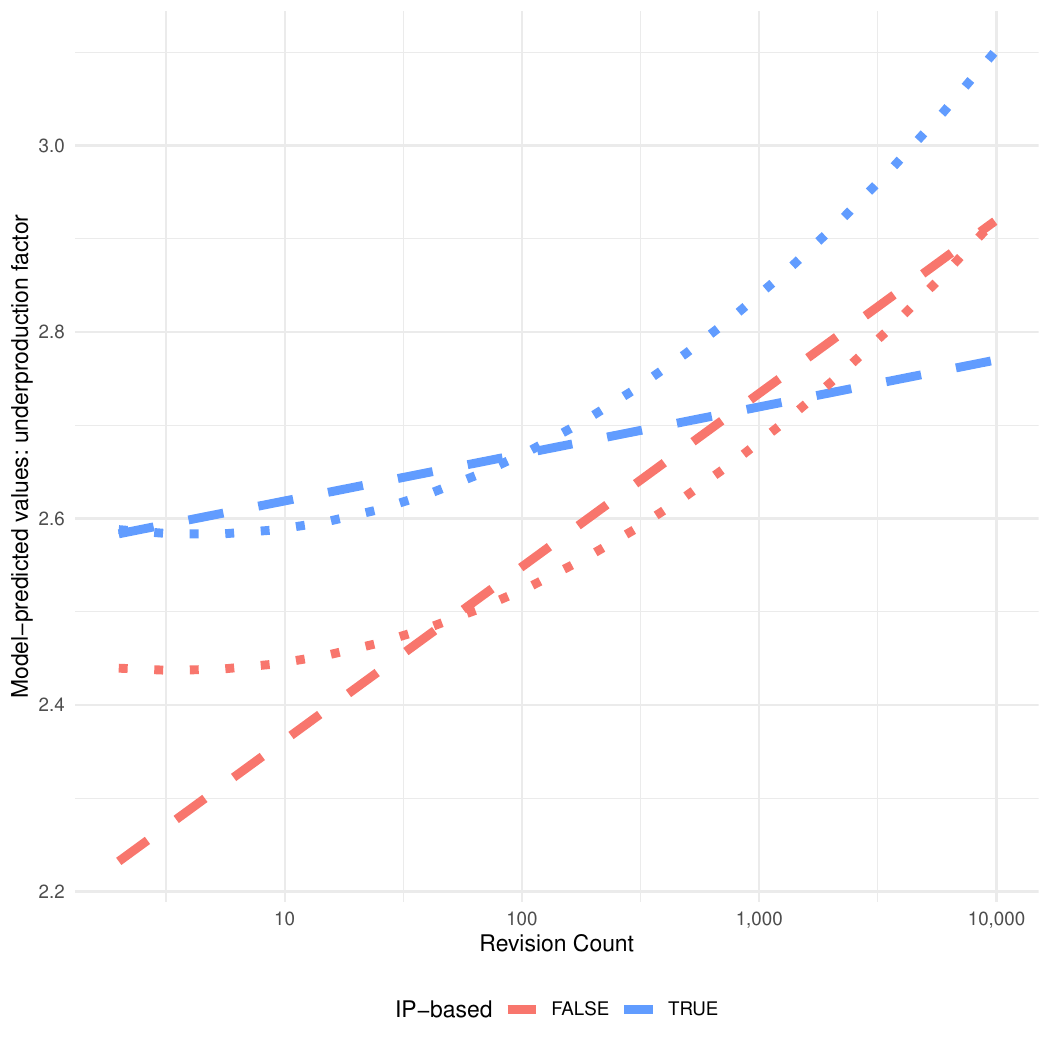}
    \caption{Marginal effect of increased experience on contributor task selection from our within-person sample. We use median individual-level fixed effects. Dashed lines are predicted values using a linear model, while dotted lines use a quadratic model, see Table \ref{tab:withinAlign}.
    \label{fig:withinMargEff}}
\end{figure}

To understand the implications of these models, we simulated task selection across experience levels, with the individual-level fixed effect set at the model estimated median. Figure \ref{fig:withinMargEff} shows the model-predicted marginal effect of accumulating editing experience on task selection. The figure shows that contributors tend towards underproduced articles as they accumulate experience. 

Overall, our results from these models provide evidence that contributors' task selection shifts toward underproduced articles as their experience increases. This trend is present in people who contribute with and without accounts. However, some models point in the opposite direction of our hypothesis in {H3B} that those who contribute without an account will shift sooner than those who use an account. We offer additional interpretation of these results in our supplement.


\section{Limitations}
\label{sec:limitations}

Although we measure the revision count of IP-based contributors, IP addresses change in ways that mean we cannot consistently associate individuals with IP addresses. Furthermore, some IP-based revisions are likely to be authored by experienced Wikipedians. As a result, our model may understate the influence of editing with an IP at low experience or overstate it at high experience.

Using revision count to operationalize experience introduces important limitations.
For example, our measure does not distinguish between a revision that fixes a typo and one that adds a paragraph. Similarly, it does not differentiate between long-lived text and quickly deleted text.
We use revision count because it is the measure of editor experience most widely used by both Wikipedians and Wikipedia researchers and because the most discussed alternative measures based on content persistence are both computationally complex and require a range of difficult decisions. Other research has shown that revision count is highly correlated with, and frequently leads to similar results as, measures that account for content persistence \citep[e.g.,][]{hill_hidden_2021}.
Ultimately, assessing how contribution types vary over the lifespan of editors remains a subject for future research.

Additionally, our design does not address causal factors among quality, popularity, and task selection. Indeed, we use measures that cannot be interpreted causally due to their granularity: our measures of quality and popularity are taken at the monthly level, not at the moment of editing. 

Another threat was described by the Wikipedia editing community when \citepos{warncke-wang_misalignment_2015} work on misalignment was first published. Community members stated that one explanation for underproduction is that some articles are more difficult to write than others and that this difficulty may be systematically distributed in a way that coincides with how people search for and consume articles.\footnote{Wikipedia community-run internal newsletter, The Signpost: \url{https://perma.cc/GX5U-32R9}} 
%
There may be factors (such as conceptual generality) driving both high popularity and low quality. In part, using ORES as a quality measure mitigates this risk because ORES measures quality based on structural characteristics independent of conceptual qualities like completeness. However, this approach does not address the fact that, as a tertiary source, Wikipedia is limited by the availability of reliable published information, including the biases in the topics these sources cover. Writing a high-quality article about a high-interest, under-reported topic may be more difficult than writing about widely documented topics.

\section{Discussion}
\label{sec:discussion}

\subsection{Born, Made, or Something Else?}

This article offers insight into community-wide trends and within-contributor changes in task selection behaviors. Our results in {H1} and {H3A} suggest that as they increase in experience, contributors tend to shift their attention to underproduced content. This is consistent with the ``made'' argument that accumulating experience is associated with editing underproduced articles and, presumably, with increased exposure to socialization. Our results in {H2} and {H3B} call this interpretation into question. Socialization suggests a social process, but contributors who participate without an account have less opportunity to build a social identity in Wikipedia. How does this occur if these contributors also shift towards underproduced articles? 

Our models suggest that IP-based contributors are initially less inclined to select underproduced articles but are comparable to account-based contributors for moderate contribution levels (i.e., in the 100-200 range).  This finding opens up a new set of challenges for social research. Given that contributors without accounts also move toward underproduced goods---and hence may represent a valuable subgroup to seek to influence and retain---how should we explain their persistence and changing task selection?

One way to understand this shift in the behavior of IP-based contributors is through an extension of learning theory---e.g.,  \citepos{preece_reader--leader_2009} reader-to-leader perspective and \citepos{lave_situated_1991} legitimate peripheral participation. As participants gain experience, perhaps their awareness of underproduced articles increases. Given that this process is mediated through technology in rich social computing environments, this may be evidence of what we call ``technosocial learning,'' rather than social learning \textit{per se}. 

We suggest that technosocial learning may be an important mechanism to shape behavior in peer production environments. For example, consider the general phenomenon of feedback on a contribution. 
Someone contributing without an account might not receive such feedback. However, this contributor could notice whether their past contribution remains present. 

Platforms offer other opportunities to learn from technical traces: contributors might experience success in mastering a particular formatting trick, observing the work of others, reading documentation, or seeing feedback others receive. 
A technosocial learning process, illustrated in Table \ref{tab:twobytwoTSLearn}, may provide a mechanism for the changes in task selection we observe.

\begin{table}
\begin{tabular}{lll}
                             & \textbf{implicit}                    & \textbf{explicit}                      \\ \cline{2-3} 
\multicolumn{1}{l|}{\textbf{social}} & \multicolumn{1}{l|}{observing the interests of others}        & \multicolumn{1}{l|}{personal feedback} \\ \cline{2-3} 
\multicolumn{1}{l|}{\textbf{technical}} & \multicolumn{1}{l|}{noticing a good is low quality} & \multicolumn{1}{l|}{automated feedback}        \\ \cline{2-3} 
\end{tabular}\caption{Dimensions of technosocial learning in peer production with examples. We might expect contributors to learn both how to contribute and about underproduction through these channels. Observing public interest and noticing low quality are implicit signals available without creating an account.\label{tab:twobytwoTSLearn}}
\end{table}

\subsection{How may underproduction be countered?}

Given the previous finding of widespread underproduction in peer production \citep{warncke-wang_misalignment_2015,champion_underproduction_2021}, countering it is an important challenge. 
One approach to addressing underproduction is to increase retention overall. 
Another approach is nudging contributors towards more severe areas of underproduction earlier: both when they are newer to platform, and more quickly in response to emerging needs. 
For example, peer production communities responded during the COVID-19 pandemic by developing FLOSS analytic and visualization tools \citep{wang_large-scale_2021}, organizing to produce Wikipedia articles to provide reliable information at high speed \citep{avieson_editors_2022}, and producing personal protective equipment and filling supply chain gaps \citep{sarkar_digital_2023}. 
Evaluation of recommender systems suggests they can influence contributors to improve underrepresented topics, providing the recommendation's relevance is kept constant \citep{houtti_leveraging_2023}.
Creating interest groups within a larger project may also be an effective way to tackle underproduction. In Wikipedia, contributors have self-organized around themes of interest to identify and work to close a wide range of participation and content gaps---e.g., WikiProject Women in Red \citep{tripodi_ms_2021}, WikiProject Women Scientists \citep{halfaker_interpolating_2017}, and WikiProject Vital Articles \citep{houtti_we_2022}.

However, more work is needed to understand when these interventions are effective. \citet{ford_beyond_2018} found that notification of existing contributors, writ large, was insufficient to address content gaps, instead recommending recruiting people with expertise related to the gap. A computational linguistics analysis from \citet{schmahl_is_2020} found that gendered bias in articles changed modestly, at best, between 2006 and 2020 when many targetted interventions were conducted. 
Successful interventions in task selection include not only involving diverse participants but also participants with diverse motives (e.g., through course assignments and non-profit outreach activities \citep[as per][]{konieczny_wikis_2012,coelho_why_2018,gallagher_teaching_2019, xing_editing_2020,herfort_spatio-temporal_2023,karczewska_glam_2023}).
\subsection{Problematizing retention and account creation}
Our analysis points to promoting contributor retention as a means of countering underproduction. Scholarship in online community and peer production community management has also emphasized retention as a path toward building healthy online communities \citep[e.g.,][]{kraut_building_2012}. However, our findings should be read alongside the evidence against requiring account creation and the risks associated with overemphasizing the retention of existing users.

Requiring accounts has been found to carry unintended consequences in terms of diminishing overall community activity levels \citep[e.g.,][]{hill_hidden_2021}. Being prompted to create an account increases the barrier to contributing. 
Maintaining a low barrier to entry may be essential to the success of peer production projects.
Beyond this risk, \citet{shaw_laboratories_2014} found that governance of peer production tends to concentrate power in the hands of elite long-standing contributors, emphasizing that longevity risks further empowering elites.
Additional analysis of long-term contributors appears in our supplement.  

Increased retention may also magnify systemic biases reflected in peer production. For example, previous work has explored geographical biases in OSM \citep{thebault-spieker_geographic_2018}, gender inequality in FLOSS participation \citep{qiu_gender_2023}, gender bias in how FLOSS projects engage with contributors \citep{chatterjee_aid_2021}, gender inequality in Wikipedia contributors \citep{hill_wikipedia_2013} and gender bias in Wikipedia content \citep{adams_wikipedia_2015}.
Therefore, while retaining contributors is valuable, it may reinforce inequality in the population of early contributors.
Scholarship has shown that communities can make progress in closing participation gaps with interventions that address bias in sociotechnical processes through which people are retained unequally \citep{karczewska_glam_2023, langrock_gender_2022, hilderbrand_engineering_2020, evans_editing_2015}. 

Along with recruiting and retaining diverse contributors, we believe there is also value in retaining those who contribute without accounts. 
A proposed update to Wikipedia's software slated for 2024 describes a change to the way that users without accounts are identified by replacing visible IP addresses with temporary identifiers.\footnote{See implementation proposal: \url{ https://perma.cc/3H59-YLH8}} 
Although details of this proposal are still in flux, this change will improve the privacy of those contributing without an account by hiding IP addresses that may reveal geographic location. Although the proposed alternative will improve the ability of researchers to track contributions from an individual across locations, it will also make it impossible to do studies of long-term engagement of users contributing without accounts---like ours.
Evidence-based management of peer production communities faces multiple critical challenges: countering underproduction, enabling diverse participation, expanding community safety, building long-term commitment, and minimizing power concentration. Some of these goals may force difficult tradeoffs, as interventions intended to address one area may make others more difficult to address. 

\section{Conclusion}
\label{sec:conclusion}

Although previous research suggests that the underproduction of peer produced goods is widespread, understanding how underproduction happens and how it can be addressed is still emerging. Our work advances a theory that experience contributes to changes in production. We test this theory and find some support: experienced contributors tend to select underproduced goods, whether they contribute with or without an account. However, contrary to our expectations, contributors not using an account are not different than those contributing using an account. In addition to the explanations of social rewards explored in previous work, we suggest that task selection changes in contributors without an account may be explained by our proposed notion of `technosocial learning.'
Our modern communication environment relies on peer production for both content and infrastructure. Understanding how to better counter underproduction is a necessary part of protecting and continuing to expand the benefit of public goods developed through peer production.


\bibliography{bibliography}
\clearpage

\section*{Acknowledgements}
Morten Warncke-Wang of the WMF provided both valuable advice on multiple occasions and access to his dataset. Aaron Halfaker provided advice in the effective use of the ORES machine learning system for quality analysis. A pilot version of this project was completed as part of the Advanced Regression course taught by Jeffrey Arnold in the Center for Statistics in the Social Sciences at University of Washington. Members of the Community Data Science Collective provided advice on early drafts of this project, including Jeremy Foote, Floor Fiers, Wm Salt Hale, Sohyeon Hwang, Sejal Khatri, Charles Kiene, Sneha Narayan, Aaron Shaw, and Nathan TeBlunthuis. 
The creation of our dataset was aided by the use of advanced computational, storage, and networking infrastructure provided by the Hyak supercomputer system at the University of Washington. 
The authors gratefully acknowledge support from the National Science Foundation, awards CNS-1703736 and CNS-1703049, and the Sloan Foundation through the Ford/Sloan Digital Infrastructure Initiative, Sloan Award 2018-11356. 
\end{document}